\documentstyle[prb,aps,epsf,multicol,epsfig]{revtex}
\begin{document}

\title{Patterned Geometries and Hydrodynamics at the Vortex Bose Glass Transition}

\author{M. Cristina Marchetti$^\dagger$ and David R. Nelson$^*$}
\address{$^*$Lyman Laboratory of Physics, Harvard University, Cambridge, MA 01238}
\address{$^\dagger$Physics Department, Syracuse University, Syracuse, NY 13244} 
\date{\today}

\maketitle

\begin{abstract}
Patterned irradiation of cuprate superconductors with columnar defects allows a new 
generation of experiments which can probe the properties of vortex liquids by confining them to 
controlled geometries. 
Here we show that an analysis of such experiments that combines an inhomogeneous
Bose glass scaling theory with the hydrodynamic description of viscous flow of vortex liquids
can be used to infer the critical behavior near the Bose glass transition.
The shear viscosity is predicted to diverge as $|T-T_{BG}|^{-z}$ at the Bose glass
transition, with $z\simeq 6$ the dynamical critical exponent.
\pacs{PACS: 70.60-w, 74.60Ec}
\end{abstract}

\begin{multicols}{2}
In the mixed state of cuprate superconductors the magnetic field is concentrated in an array of 
flexible flux bundles that, much like ordinary matter, can form crystalline, liquid
and glassy phases\cite{crabtree_nelson}.  The dynamics of the flux-line array determines the resistive 
properties of the material and has therefore been the focus of much attention. 
Novel types of glasses are also possible because of pinning in 
disordered samples \cite{mpaf}.
In particular,
the introduction of columnar damage tracks by heavy-ion irradiation
yields a low-temperature ``Bose glass'' phase, in which every vortex is trapped on a
columnar defect \cite{drnvv} and the pinning efficiency of vortex lines 
is strongly enhanced \cite{budhani,koncz,civale}. At high temperatures the vortices delocalize in an entangled 
flux-line liquid. The high temperature liquid transforms into a Bose glass via a second order 
phase transition at $T_{BG}$, characterized by universal critical exponents \cite{drnvv,bariloche}.
We show here that there are very strong divergences in the vortex shear viscosity and other  
transport coefficients as this transition is approached from the liquid, similar to behavior 
conjectured for glass transitions in ordinary forms of matter, and propose experiments which test 
our predictions. Vortex matter with columnar defects thus provides a concrete example of a 
glassy phase accessed via a genuine second order phase transition and characterized by universal critical exponents.

The Bose glass transition has been studied theoretically by viewing the vortex line trajectories as the world
lines of two-dimensional quantum mechanical particles \cite{nelsonprl,drnvv}. The thickness of the superconducting
sample corresponds to the inverse temperature of the ficticious quantum particles. In thick samples the physics
of vortex lines pinned by columnar defects becomes equivalent to the low temperature properties
of two-dimensional bosons with point disorder. The low temperature phase is a Bose glass 
where the vortices behave like localized bosons. It has vanishing linear resistivity and an infinite tilt
modulus \cite{drnvv}. The entangled flux liquid phase is resistive and corresponds to a boson superfluid
\cite{matching}.

Although an exact theory of the continuous transition at $T_{BG}(B)$ from the Bose glass to the 
entangled flux liquid (or ``superfluid'')
is not available, most physical properties can be described via a scaling theory in terms of just 
two undetermined critical exponents \cite{drnvv,drnleo,fwgf}.
In the low temperature Bose glass each flux line is localized in the vicinity of one or more 
columnar pins. Its excursion in the direction perpendicular to the applied field is 
characterized by a correlation length that diverges at $T_{BG}$,
$l_\perp(T)\sim|T-T_{BG}|^{-\nu_\perp}$.
There is also a diverging correlation length along the applied field direction (here the $z$ direction),
$l_\parallel(T)\sim|T-T_{BG}|^{-\nu_\parallel}$, where 
$\nu_\parallel=2\nu_\perp$ \cite{fwgf}.
The time scale $\tau$ for relaxation of a fluctuation of size $l_\perp$ is assumed to diverge with
a critical exponent $z$,
$\tau\sim l_\perp^z\sim|T-T_{BG}|^{-z\nu_\perp}$\cite{drnvv}.
The universal critical exponents as determined by the most recent simulations are
$\nu_\perp\simeq 1$ and $z\simeq 4.6\pm 2$ \cite{wallin}.
Scaling can then be used to relate physical quantities to these diverging 
length and time scales. In particular, the  resistivity $\rho(T)$ for currents applied in the $ab$ plane
is predicted to vanish as $T\rightarrow T_{BG}$ from above as
$\rho\sim |T-T_{BG}|^{\nu_\perp(z-2)}$ \cite{drnvv}.
Some predictions of the scaling theory  have been tested experimentally, but 
there are as yet no direct measurements of the transport coefficients usually associated
with glass transitions in conventional forms of matter, such as the shear viscosity.
As we shall see, the behavior of the shear viscosity is determined by
the dynamical critical exponent $z$ that controls the divergence of the relaxation
time in the Bose glass phase. A measurement of the shear viscosity
would provide a direct probe of the diverging relaxation time
associated with glassy behavior \cite{kes}.

Patterned irradiation of cuprate superconductors 
with columnar defects allows for a new 
generation of experiments that may in fact provide a direct probe of viscous
critical behavior
near the Bose glass transition \cite{pastoriza}. By starting with a clean sample,
at temperatures such that point disorder is negligible,
it is possible to selectively  irradiate  regions
of controlled geometry. An example is shown in Fig. 1.
The side regions have been heavily irradiated, and are characterized by a high matching field
$B_{\phi}^{(2)}$ and transition curve $T_{BG}^{(2)}$, 
while the channel is lightly irradiated with a lower matching field
$B_{\phi}^{(1)}<B_{\phi}^{(2)}$ and transition curve $T_{BG}^{(1)}$. 
When $T_{BG}^{(1)}<T_{BG}<T_{BG}^{(2)}$, the flux array in the channel
is in the liquid state, while the contacts are in the Bose glass phase.
Flow in the resistive
flux liquid region is impeded by the ``Bose-glass contacts'' at the boundaries, as the many trapped vortices in
these regions provide an essentially impenetrable barrier for the flowing vortices. 
As discussed in Ref. \onlinecite{mcmdrn}, the pinning at the boundaries propagates into the liquid channels by
a viscous length $\delta$ that depends on the flux liquid viscosity. 
As the temperature is lowered at constant field, so that
the Bose glass transition $T_{BG}^{(1)}$ of the liquid region is approached from above
(Fig. 2) the growing Bose glass
correlations increase $\delta$ and strongly suppress  the flow in the channel and the associated flux
flow voltage drop across the channel.
\begin{figure}[htb]
\epsfxsize=3.in
\epsfbox{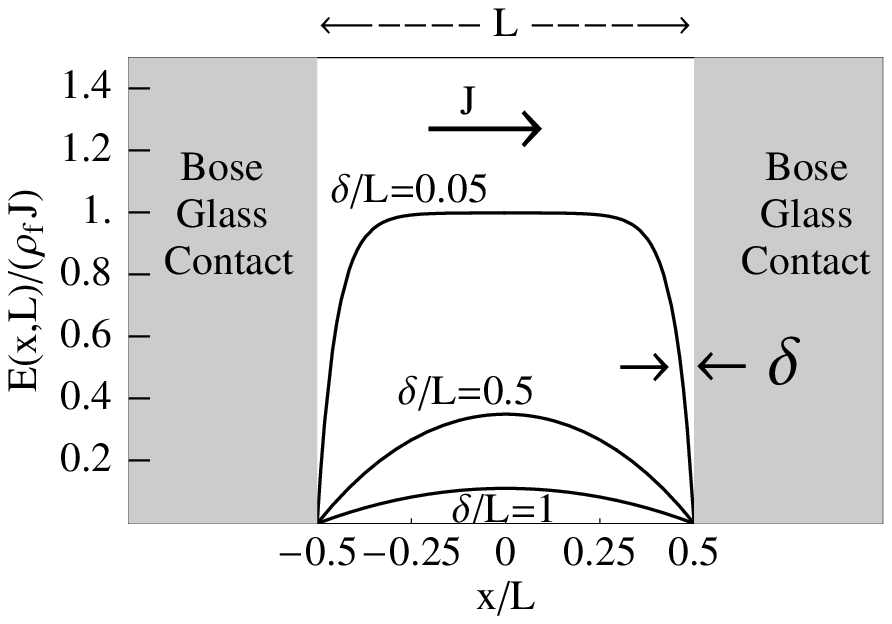}
\label{channel}

FIG. 1. A weakly irradiated channel (white region) where the flux liquid is sandwiched between 
two heavily irradiated Bose-glass contacts (shaded regions). 
A current $J$ applied across the channel yields flux motion
along the channel. The reduced field profiles given by Eq. (9) are shown
for a few values of $\delta/L$ and can be measured by a series of voltage taps. 
\end{figure}

In this paper we analyze experiments with flux flow
in such confined geometries by combining  the predictions 
of the Bose glass scaling theory {--} generalized to the spatially inhomogeneous case {--}
with the hydrodynamics 
of viscous flow of vortex liquids \cite{mcmdrn}. Our analysis shows that
the viscous length $\delta$ controlling boundary pinning is just the
Bose-glass localization length, $l_\perp$, and therefore provides a prescription for measuring the 
Bose glass scaling
near the transition. Both flow in the channel geometry sketched in Fig. 1 and in the Corbino disk
geometry (Fig. 3) used recently by L\'opez et al. \cite{lopez} is discussed. Such experiments can be used 
to extract the critical behavior of various 
transport coefficients and map out the entire critical region. In particular,
the flux liquid shear viscosity is predicted to diverge as $|T-T_{BG}|^{-z}$ at the Bose glass
transition. Because $z\simeq 4.6\pm 2.0$, this powerful divergence is reminiscent
of the Vogel-Fulcher behavior $\eta\sim\exp\big[c/(T-T_g)\big]$ conjectured
for glass transitions in conventional forms of matter.


The Bose glass scaling theory summarized earlier is easily generalized to the case of
spatially inhomogeneous flow in constrained geometries. Considering for simplicity the channel geometry,
a generalized scaling ansatz for the local electric field from flux motion at position $x$ in a channel of
thickness $L$ takes the form
\begin{equation}
\label{scaling}
E(T,J,x,L)=b^{-(1+z)}E\bigg(b^{1/\nu_\perp}t,{b^{\nu_\perp}b^{\nu_\parallel}J\phi_0\over ck_BT},
{x\over b},{L\over b}\bigg),
\end{equation}
where $b>1$ is the length scaling parameter and $t=|T-T_{BG}|/T_{BG}$  the reduced temperature. 
This ansatz follows from the usual assumption that the continuous transition is described by
a single diverging length scale and the  homogeneity condition on the relevant physical quantities
at the transition
(see, e.g.,  Refs. \onlinecite{drnvv,fwgf}). 
\begin{figure}[bth]
\epsfxsize=3.0truein
\epsfbox{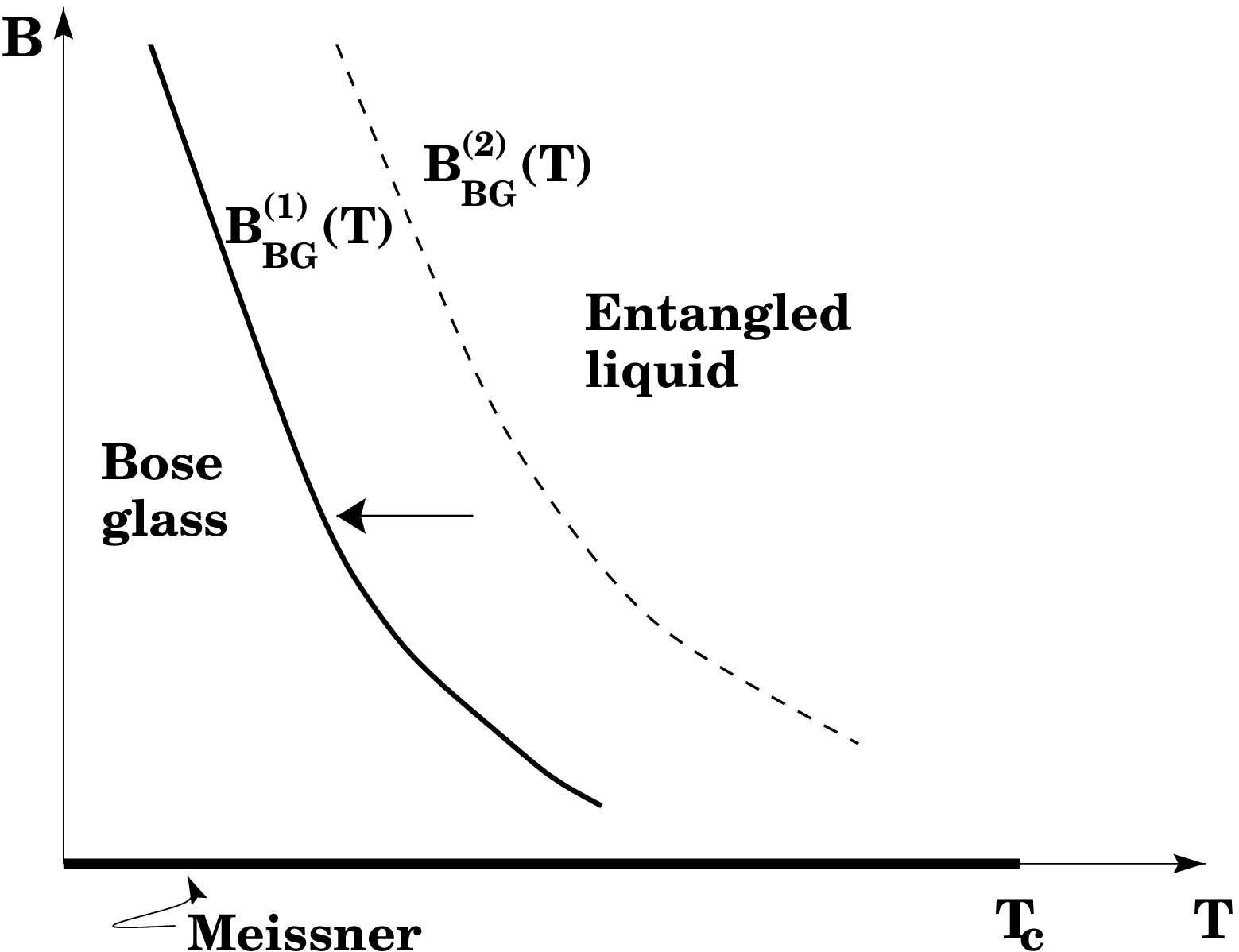}
\label{PhaseDiag}

FIG. 2. A sketch of the $(B,T)$ phase diagram for the flux array in the weakly irradiated channel region.
The heavy line $B_{BG}^{(1)}(T)$ denotes the continuous transition from the Bose glass to the entangled liquid.
Also shown is the location $B_{BG}^{(2)}(T)$ of the Bose glass transition line in the heavily irradiated 
contacts. When a field $B_{BG}^{(1)}(T)<B<B_{BG}^{(2)}(T)$ is applied, the flux array in the channel is in the liquid state,
while the contacts are in the Bose glass phase. By decreasing the temperature at constant field as
indicated by the arrow, the Bose glass transition of the channel region is approached from above. 
\end{figure}
The response in the
Bose glass is generally nonlinear in the applied current $J$.
By choosing $b=t^{-\nu_\perp}\sim l_\perp(T)$ we obtain
\begin{equation}
\label{scaling2}
E(T,J,x,L)=l_\perp^{-(1+z)}E\Big(1,{l_\perp l_\parallel J\phi_0\over ck_BT},{x\over l_\perp},{L\over l_\perp}\Big).
\end{equation}
In the entangled flux liquid the response is linear at small current.
Upon expanding the right hand side of Eq. (\ref{scaling2}) we obtain for $J\rightarrow 0$
\begin{equation}
\label{linear_scaling}
E(J\rightarrow 0,x,L)\simeq \rho_0 \Big({l_\perp\over a_0}\Big)^{2-z} J {\cal F}(x/l_\perp,L/l_\perp),
\end{equation}
where $a_0$ is the vortex spacing and $\rho_0=\Big(n_0\phi_0/c\Big)^2(1/\gamma_0)$ is the Bardeen-Stephen resistivity of noninteracting
flux lines, with $\gamma_0$ a bare friction. 
A scaling form for the resistivity $\rho(T,L)=\Delta V/(LJ)$, with $\Delta V$ the net voltage drop across the channel,
is easily obtained by integrating Eq. (\ref{linear_scaling}), with the result,
\begin{equation}
\label{rho}
\rho(T,L)=\rho_f(T)f(L/l_\perp)
\end{equation}
with $f(x)={1\over x}\int_0^x du{\cal F}(u,x)$ a scaling function and $\rho_f(T)$
the bulk resistivity,
\begin{equation}
\label{rhobulk}
\rho_f(T)=\rho_0\Big({l_\perp\over a_0}\Big)^{2-z}
\equiv\Big({n_0\phi_0\over c}\Big)^2{1\over\gamma}.
\end{equation}
In the second line of Eq. (\ref{rhobulk}) the dependence on the Bose glass correlation
length $l_\perp$ has been incorporated in a renormalized  friction coefficient 
$\gamma=\gamma_0\Big({l_\perp\over a_0}\Big)^{z-2}$
that diverges at the transition as $\gamma\sim|T-T_{BG}|^{\nu_\perp(z-2)}$ \cite{drnvv}.
For $L\gg l_\perp$, the channel geometry has no effect and one must recover the bulk result,
leading to $f(x\gg 1)\sim 1$.

The scaling function ${\cal F}$ can be determined by  assuming that
the long wavelength electric field of Eq. (\ref{linear_scaling})
is described
by hydrodynamic equations \cite{mcmdrn}.
For simple geometries where the current is applied in the $ab$ plane and the flow is spatially
homogeneous in the $z$ direction, these reduce to a single equation 
for the coarse-grained flux liquid flow velocity ${\bf v}({\bf r})$, \cite{hall}
\begin{equation}
\label{hydro}
-\gamma{\bf v}+\eta\nabla^2_\perp{\bf v}+{\bf f}_L=0,
\end{equation}
The second term in Eq. (\ref{hydro}) is the flux liquid viscosity $\eta(T,H)$ and
represents the viscous drag arising
from intervortex interactions and entanglement. 
Finally, ${\bf f}_L=-{1\over c}n_0\phi_0\hat{\bf z}\times{\bf J}$ is the Lorentz force density driving
the flux motion.
Intervortex interaction at the Bose-glass boundaries translates into a no-slip boundary
condition for the flux liquid flow velocity. By preventing the free flow of flux liquid,
the Bose glass boundaries can significantly decrease the macroscopic flux-flow resistivity of
the superconductor.
Once the velocity field is obtained by solving Eq. (\ref{hydro}) with suitable boundary conditions,
the electric field profile in the superconductor is found immediately from
${\bf E}({\bf r})={n_0\phi_0\over c}\hat{\bf z}\times{\bf v}({\bf r})$.
It is instructive to rewrite Eq. (\ref{hydro}) as an equation for the local electric field,
\begin{equation}
\label{viscousE}
-\delta^2\nabla^2_\perp{\bf E}+{\bf E}=\rho_f{\bf J},
\end{equation}
where $\delta=\sqrt{\eta/\gamma}$ is the viscous length. When the first term on the right hand
side is absent, i.e., the flux liquid viscosity is small, this equation of ``viscous electricity''
reduces to Ohm's law with flux flow resistivity given by the bulk value,
$\rho_f(T)$. Interactions, however, make the viscous drag important
and as a result  the electrodynamics of flux-line liquids is highly nonlocal
near the Bose glass transition.

The solution of the hydrodynamic equation 
for the simple channel geometry sketched in Fig. 1, with a homogeneous current
${\bf J}=-\hat{\bf x} J$ applied across the channel,
is given by
\begin{equation}
\label{Echannel}
E(x,L)=\rho_f J \bigg[1-{\cosh(x/\delta)\over \cosh(L/2\delta)}\bigg],
\end{equation}
and is shown in Fig. 1.
Upon  comparing Eq. (\ref{Echannel}) to Eq. (\ref{linear_scaling}), we see that the quantity in square
brackets in Eq. (\ref{Echannel}) is the scaling function ${\cal F}$ and find that the viscous length $\delta$ 
is in fact the Bose glass length $l_\perp$. As the friction diverges at $T_{BG}$ according to
$\gamma\sim |T-T_{BG}|^{-\nu_\perp(z-2)}$,
this identification immediately gives that the flux liquid shear viscosity also diverges at the
Bose glass transition with
\begin{equation}
\eta=l_\perp^2\gamma\sim |T-T_{BG}|^{-\nu_\perp z}.
\end{equation}
The scaling form for the resistivity is obtained by
integrating Eq. (\ref{Echannel}), with the result
\begin{equation}
\rho(T,L)=\rho_f(T)\Big[1-{2l_\perp\over L}\tanh\Big({L\over 2l_\perp}\Big)
\Big].
\end{equation}
If $l_\perp\ll L$, we recover the bulk result of Eq. (\ref{rhobulk}),
$\rho(T,L)=\rho_f(T)\sim |T-T_{BG}|^{\nu_\perp(z-2)}$.
Near the transition, where $l_\perp\gg L$, the resistivity depends on the channel width and is
controlled by the shear viscosity, with
\begin{equation}
\rho(T,L)\simeq {\rho_fL^2\over 12 l_\perp^2}=
\Big({n_0\phi_0\over c}\Big)^2{L^2\over 12\eta(T)}\sim L^2|T-T_{BG}|^{\nu_\perp z}.
\end{equation}
This strong divergence of the viscosity is precisely the kind of behavior
expected at a liquid-glass transition. In this sense the Bose glass transition 
is an example of a glass transition that is well understood theoretically and 
where precise predictions are available.

\begin{figure}[bth]
\epsfxsize=3.0in
\epsfbox{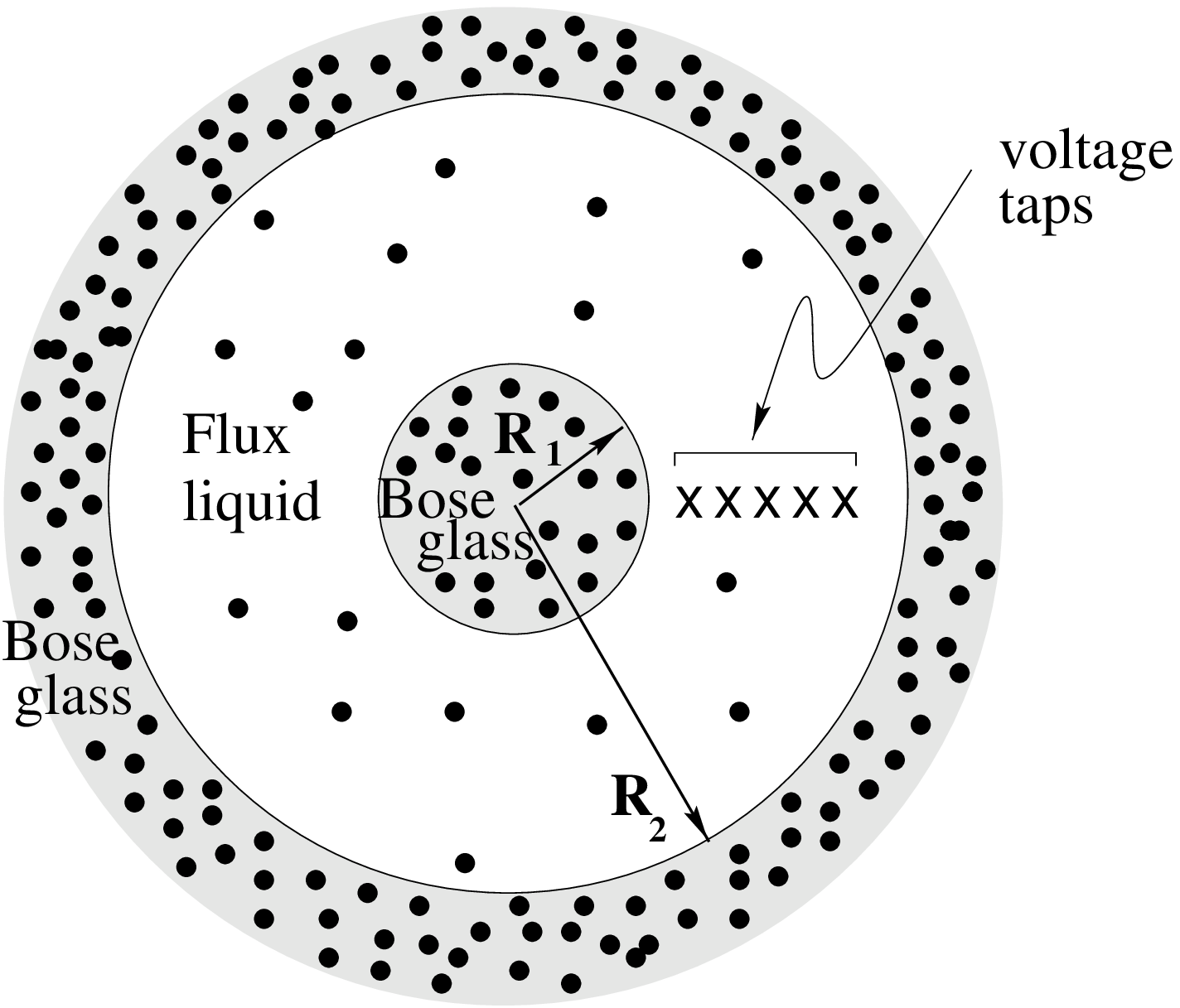}
\label{corbino_topview}

FIG. 3. Top view of the Corbino disk geometry with Bose glass contacts. The magnetic field is
out of the page. The vortex array is in 
the Bose glass state in the inner and outer densely dotted regions and in the flux liquid state 
in the weakly irradiated annular region. A radial driving current drives flux motion in the azimuthal direction,
and the voltage taps allow the electric field profile to be determined.
\end{figure}

Another important patterned geometry is the Corbino disk, recently used by L\'opez et al. for defect-free materials \cite{lopez} to reduce boundary effects in the flux flow measurements.
Here we propose fabrication of a Corbino disk with Bose glass inner and outer contacts sketched in Fig. 3.
A current $I$ injected at the outer boundary and extracted at the inner boundary yields a radial 
current density ${\bf J}(r)=-{I\over 2\pi(R_2-R_1)}{\hat{\bf r}\over r}$ that drives vortex motion in the azimuthal direction.
The electric field induced by flux motion is radial, ${\bf E}({\bf r})=-E(r)\hat{\bf r}$, and 
its magnitude is obtained by solving
Eq. (\ref{viscousE}) in a cylindrical geometry, with the result,
\begin{equation}
\label{efieldmag}
E(r)={\rho_f I\over 2\pi (R_2-R_1)l_\perp}\bigg[{l_\perp\over r}+c_1I_1({r\over l_\perp})
   +c_2K_1({r\over l_\perp})\bigg],
\end{equation}
where
\begin{eqnarray}
& & c_1={K_1(\rho_2)/\rho_1-K_1(\rho_1)/\rho_2\over
        K_1(\rho_1)I_1(\rho_2)-K_1(\rho_2)I_1(\rho_1)} \\ \nonumber
& & c_2={I_1(\rho_1)/\rho_2-I_1(\rho_2)/\rho_1\over
        K_1(\rho_1)I_1(\rho_2)-K_1(\rho_2)I_1(\rho_1)},
\end{eqnarray}
with $\rho_{1,2}=R_{1,2}/l_\perp$ and $I_1(x)$ and 
$K_1(x)$ Bessel functions. 
The electric field profiles are shown in Fig. 4.
The resistivity is defined in terms of the net voltage drop $\Delta V_{12}$
between the inner and outer radii as $\rho(T,R_1,R_2)=\Delta V_{12}/[I/2\pi(R_2-R_1)]$.
Near the Bose glass transition, where $l_\perp\gg R_2, R_1$, we find
\begin{equation}
\label{rhocorbino}
\rho\simeq{(n_0\phi_0/ c)^2\over 4\eta(T)}
  \Big\{{R_2^2-R_1^2\over 2}-{4R_1^2R_2^2[\ln(R_2/R_1)]^2\over R_2^2-R_1^2}\Big\}.
\end{equation}
As in the channel geometry, the resistivity at the transition is completely determined by the diverging
viscosity and the geometrical parameters of the channel.
\begin{figure}[bth]
\epsfxsize=3.in
\epsfbox{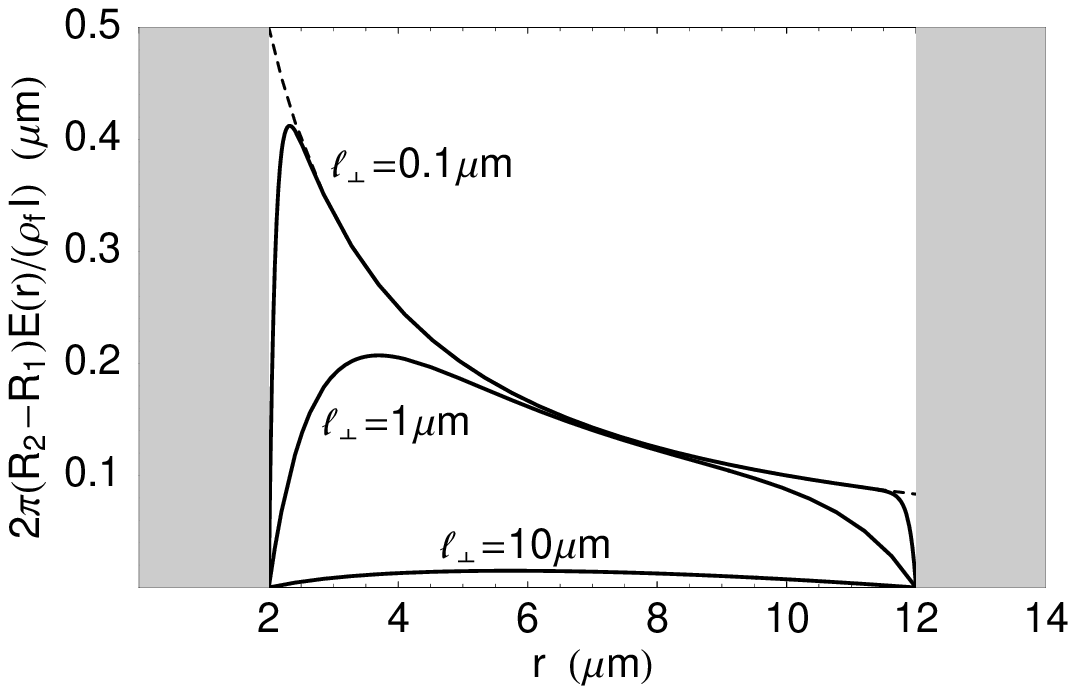}
\label{CorbinoProfiles}

FIG. 4. The electric field profile for the Corbino 
disk geometry (Eq. (\ref{efieldmag})), with
$R_1=2\mu m$ and $R_2=12\mu m$, with a width $R_2-R_1=10\mu m$.
The vertical axis represents the reduced field $2\pi E(r)/(\rho_f I)$
in $\mu m$. 
The dashed line is the electric field  for vanishing viscosity,
$E_0(r)={\rho_f I\over 2\pi(R_2-R_1)}{1\over r}$.
The shaded regions represent the Bose glass contacts.
\end{figure}
Experiments with patterned geometries near the Bose glass transformation provide
an exciting opportunity to probe viscous behavior near a second order glass transition.
A similar scaling analysis leads to predictions for the additional viscosities which characterize the 
dynamics of vortex matter \cite{mcmdrn,huse}. For example, the viscous generalization
of Ohm's law for transport parallel
to the applied field reads
\begin{equation}
\label{Eparallel}
-\delta_\parallel^2 \partial^2_zE_\parallel -\delta^2_\perp \nabla^2_\perp E_\parallel
+E_{\parallel}=\rho_\parallel J_\parallel,
\end{equation}
with $\rho_\parallel\sim l_\perp^{-z}$, $\delta_\parallel\sim l_\parallel$, and 
$\delta_\perp\sim l_\perp$. 

\vskip 0.2in
This work was supported by the National Science Foundation at Syracuse through Grants No. DMR97-30678 and DMR98-05818
and at Harvard through Grant No. DMR97-14725, and by the Harvard Materials Research Science and Engineering Center
through Grant No. DMR98-09363.

\end{multicols}


\begin{references}
\bibitem{matching}
For simplicity, we confine our attention here to fields less that the ``matching field'' 
$B_\phi=n_{\rm pin}\phi_0$,
with $n_{\rm pin}$ the areal density of columnar pins and $\phi_0=\hbar c/2e$ the flux quantum.
\bibitem{hall}
For simplicity we have neglected here the small component of the frictional force 
on the flux liquid normal to the flow velocity, which is responsible for the Hall effect.
\end{references}
\end{document}